\documentclass[sigconf]{acmart}

\usepackage{graphicx}
\usepackage{latexsym}
\usepackage{url}            
\usepackage{booktabs}       
\usepackage{nicefrac}       
\usepackage{latexsym}
\usepackage{xspace}
\usepackage{multirow}
\usepackage{subcaption}
\usepackage{amsmath}
\usepackage{wrapfig,lipsum}
\usepackage{csquotes}
\usepackage{fancyvrb}
\usepackage[export]{adjustbox}
\usepackage{array}
\usepackage{arydshln}

\newcolumntype{L}[1]{>{\raggedright\let\newline\\\arraybackslash\hspace{0pt}}m{#1}}
\newcolumntype{C}[1]{>{\centering\let\newline\\\arraybackslash\hspace{0pt}}m{#1}}
\newcolumntype{R}[1]{>{\raggedleft\let\newline\\\arraybackslash\hspace{0pt}}m{#1}}

\makeatletter
\def\adl@drawiv#1#2#3{%
        \hskip.5\tabcolsep
        \xleaders#3{#2.5\@tempdimb #1{1}#2.5\@tempdimb}%
                #2\z@ plus1fil minus1fil\relax
        \hskip.5\tabcolsep}
\newcommand{\cdashlinelr}[1]{%
  \noalign{\vskip\aboverulesep
           \global\let\@dashdrawstore\adl@draw
           \global\let\adl@draw\adl@drawiv}
  \cdashline{#1}
  \noalign{\global\let\adl@draw\@dashdrawstore
           \vskip\belowrulesep}}
\makeatother

\newcommand{\matchpyramid}{\text{MP}\xspace}
\newcommand{\prf}{\text{RM3 PRF}\xspace}
\newcommand{\convknrm}{\text{ConvKNRM}\xspace}
\newcommand{\pacrr}{\text{PACRR}\xspace}
\newcommand{\knrm}{\text{KNRM}\xspace}
\newcommand{\tk}{\text{TK}\xspace}

\newcommand{\coll}{\texttt{TripClick}\xspace}
\newcommand{\collurl}{\url{https://tripdatabase.github.io/tripclick}\xspace}

\newcommand{\head}{\texttt{HEAD}\xspace}
\newcommand{\torso}{\texttt{TORSO}\xspace}
\newcommand{\tail}{\texttt{TAIL}\xspace}

\newcommand{\raw}{\texttt{RAW}\xspace}
\newcommand{\dctr}{\texttt{DCTR}\xspace}
\usepackage{bm}

\def\eqref#1{equation~\ref{#1}}

\def\1{\bm{1}}

\DeclareMathAlphabet{\mathsfit}{\encodingdefault}{\sfdefault}{m}{sl}
\SetMathAlphabet{\mathsfit}{bold}{\encodingdefault}{\sfdefault}{bx}{n}

\copyrightyear{2021} 
\acmYear{2021} 
\setcopyright{acmcopyright}\acmConference[SIGIR '21]{Proceedings of the 44th International ACM SIGIR Conference on Research and Development in Information Retrieval}{July 11--15, 2021}{Virtual Event, Canada}
\acmBooktitle{Proceedings of the 44th International ACM SIGIR Conference on Research and Development in Information Retrieval (SIGIR '21), July 11--15, 2021, Virtual Event, Canada}
\acmPrice{15.00}
\acmDOI{10.1145/3404835.3463242}
\acmISBN{978-1-4503-8037-9/21/07}

\begin{document}
\fancyhead{}

\title{TripClick: The Log Files of a Large Health Web Search Engine}

\author{Navid Rekabsaz}
\email{navid.rekabsaz@jku.at}
\affiliation{%
  \institution{Johannes Kepler University Linz}
  \institution{Linz Institute of Technology, AI Lab}
  \country{Austria}
}

\author{Oleg Lesota}
\email{oleg.lesota@jku.at}
\affiliation{%
  \institution{Johannes Kepler University Linz}
  \institution{Linz Institute of Technology, AI Lab}
  \country{Austria}
}

\author{Markus Schedl}
\email{markus.schedl@jku.at}
\affiliation{%
  \institution{Johannes Kepler University Linz}
  \institution{Linz Institute of Technology, AI Lab}
  \country{Austria}
}

\author{Jon Brassey}
\email{jon.brassey@tripdatabase.com}
\affiliation{%
  \institution{Trip Database}
  \country{United Kingdom}
}

\author{Carsten Eickhoff}
\email{carsten@brown.edu}
\affiliation{%
  \institution{Brown University}
  \country{United States}
}

\renewcommand{\shortauthors}{Rekabsaz et al.}

\definecolor{mint}{rgb}{0.24, 0.71, 0.54}
\newcommand{\carsten}[1]{\textit{\textcolor{mint}{#1}}}

\begin{abstract}
Click logs are valuable resources for a variety of information retrieval (IR) tasks. This includes query understanding/analysis, as well as learning effective IR models particularly when the models require large amounts of training data. We release a large-scale domain-specific dataset of click logs, obtained from user interactions of the Trip Database health web search engine. Our click log dataset comprises approximately 5.2 million user interactions collected between 2013 and 2020. We use this dataset to create a standard IR evaluation benchmark --\coll -- with around 700,000 unique free-text queries and 1.3 million pairs of query-document relevance signals, whose relevance is estimated by two click-through models. As such, the collection is one of the few datasets offering the necessary data richness and scale to train neural IR models with a large amount of parameters, and notably the first in the health domain. Using \coll, we conduct experiments to evaluate a variety of IR models, showing the benefits of exploiting this data to train neural architectures. In particular, the evaluation results show that the best performing neural IR model significantly improves the performance by a large margin relative to classical IR models, especially for more frequent queries.
\vspace{-0.2cm}
\end{abstract}

\ccsdesc[500]{Information systems~Test collections\vspace{-0.2cm}}
\ccsdesc[500]{Information systems~Learning to rank}

\keywords{click logs, collection, health information retrieval, medical information retrieval, neural ranking models \vspace{-0.1cm}}


\maketitle

\section{Introduction}
User interactions with information systems are a valuable resource for retrieval system training, refinement and evaluation. These interactions, in the form of click logs, contain submitted queries alongside clicked documents from the result page. To be effective, these collections are sizable, and can be exploited for search engine effectiveness improvement~\cite{craswell2020overview,dai2018convolutional,xiong2017end}, as well as studying user behavior~\cite{pass2006picture}, and information needs~\cite{huang2013learning}.

In the health domain, information needs are often diagnostic, therapeutic or educational in nature. Common queries reflect patient characteristics such as demographics, general disposition or symptoms~\cite{haynes2005optimal,kuhn2016implicit,roberts2015overview,wei2018distant,wei2018embedding} and aim at obtaining a differential diagnosis~\cite{eickhoff2019diagnostic,li2020mining}, suggested treatments~\cite{haynes2005optimal}, or tests that might help narrow down the range of candidate diagnoses. In comparison with general-purpose search engines, the user base of health search engines is almost exclusively composed of domain experts (healthcare professionals) and behavioral traces may differ significantly from those observed on the popular web. 

This work develops and shares \coll, a large-scale dataset of the click logs provided by \url{https://www.tripdatabase.com}, a health web search engine for retrieving clinical research evidences, used almost exclusively by health professionals. The dataset consists of $5.2$ million clicks collected between 2013 and 2020, and is publicly available for research purposes. Each log entry contains an identifier for the ongoing search session, the submitted query, the list of retrieved documents, and information on the clicked document. \coll is one of the very few datasets providing the necessary data richness and scale to train deep learning-based IR models with a high number of parameters. To the best of our knowledge, this is the first effort to release a \textit{large-scale click log dataset in the health domain}. It can serve various information processing scenarios, such as retrieval evaluation, query analysis, and user behavior studies. In particular, covering the search activities throughout the year 2020, the \coll dataset provides an interesting resource capturing the COVID-19 pandemic. 

Based on the click logs, we create and provide a \textit{health IR benchmark}. The benchmark consists of a collection of documents, a set of queries, and the query-document relevance information, extracted from user interactions. Regarding the documents collection, since the vast majority of the retrieved and clicked documents in the dataset are medical articles originating from the MEDLINE catalog.\footnote{\url{https://pubmed.ncbi.nlm.nih.gov}} We create the IR benchmark using the subset of the click logs containing the documents in MEDLINE. This results in $1.5$ million medical articles' abstracts, $692,\!000$ unique queries, and $4$ million pairs of interactions between these queries and documents. We create and provide two estimations of query-document relevance using two click-through models~\cite{chuklin2015click}. The first one, referred to as \raw, follows a simple approach by considering every clicked document relevant to its corresponding query. The second uses the Document Click-Through Rate (\dctr)~\cite{craswell2008experimental}, which estimates query-document relevance as the rate of clicking the document over all retrieved results of a specific query. 

The \coll benchmark provides three groups of queries for evaluation of IR models. The groups are created according to specific query frequency ranges. Concretely, the \head group consists of most frequent queries which appear more than 44 times, non-frequent ones with frequencies between 6 and 44 times are grouped in \torso, and \tail encompasses rare queries appearing less than 6 times. To facilitate research on neural IR models, we create a large training set in pairwise learning-to-rank format~\cite{liu2010learning}. Each item in the training data consists of a query, one of its relevant documents, and a randomly selected non-relevant document. 

Using this data, we study the performance of several recent neural IR models as well as strong classical baselines. Evaluation is carried out using standard IR evaluation metrics, namely Mean Reciprocal Rank (MRR), Recall at cut-off 10, and Normalized Discounted Cumulative Gain (NDCG) at cut-off 10. The results show significant improvements of neural architectures over classical models in all three groups. This improvement is particularly prominent for more frequent queries, \textit{i.e.}, the ones in the \head and \torso groups.

The contribution of this work is three-fold: 
\begin{itemize}
    \item Releasing a large-scale dataset of click logs in the health domain. 
    \item Creating a novel health IR benchmark, suited for deep learning-based IR models.
    \item Conducting evaluation experiments on various classical and neural IR models on the collection.
\end{itemize}

The click logs dataset, the benchmark, and all related resource as well as the code used to create the benchmark are available on \textbf{{\collurl}}.

The remainder of this paper is structured as follows: Related resources are reviewed in Section~\ref{sec:related}. Section~\ref{sec:dataset} describes the dataset of click logs, followed by explaining the process of creating the \coll IR benchmark in Section~\ref{sec:benchmark}. We lay out our experiment setup and report and discuss the results in Section~\ref{sec:experiments}.

\section{Related Resources}
In this section, we review some of the existing resources related to \coll, in particular large-scale search log datasets in the web domain, as well as some common health IR collections. The statistics of these resources as well as our novel \coll dataset are summarized in Table~\ref{tbl:coll_stats}.

\begin{table}[t]
\begin{center}
\caption{Number of queries and number of query-document interactions (Q-D) of various IR collections in the web and health domain.}
\begin{tabular}{l l c c}
\toprule
\multicolumn{2}{c}{Collection} & \multicolumn{1}{c}{Queries} & \multicolumn{1}{c}{Q-D} \\\midrule

\multirow{4}{*}{\rotatebox[origin=c]{90}{Web}} & Sogou-QCL~\cite{zheng2018sogou} & 537K & 12.2M\\
& MS MARCO Passage Retrieval~\cite{nguyen2016ms} & 1.0M & 532K \\
& MS MARCO Document Retrieval~\cite{nguyen2016ms} & 367K & 384K  \\
& ORCAS~\cite{craswell2020orcas} & 10.4M & 18.8M\\\cdashlinelr{2-4}
\multirow{3}{*}{\rotatebox[origin=c]{90}{Health}} & \coll Logs Dataset & 1.6M & 5.2M \\
& TREC Precision Medicine 2019~\cite{Roberts2019OverviewOT} & 40 & 13K \\
& CLEF Consumer Health Search 2018~\cite{jimmy2018overview} & 50 & 26K\\
\bottomrule
\end{tabular}
\label{tbl:coll_stats} 
\end{center}
\end{table}

\begin{figure}
    \centering
\noindent\fbox{%
\small
    \parbox{0.45\textwidth}{%
\texttt{"DateCreated": Date(1510099598753)\\
SessionId: 0voniyqiiinv41t3y2jwosx0\\
Keywords: "risk of cancer from diagnostic x-rays"\\
Documents: [1184559, 9261540, 4780587, 1412562, 5002174, 5026261, 5569939, 9416551, 9410485, 5611210, 6659224, 1172157, 9279530, 4974766, 5857055, 1314398, 7875167, 1400849, 7622126, 9280769]\\
DocumentId: 6659224\\
Url: "http://www.ncbi.nlm.nih.gov/pubmed/20602108"\\ 
Title: "Diagnostic X-ray examinations and increased chromosome translocations: evidence from three studies\\
DOI: "10.1007/s00411-010-0307-z"\\
ClinicalAreas: "Radiology"
}
}}
\caption{Sample click log entry.}
\label{fig:logs}
\end{figure}

Large-scale click log datasets in the English Web domain have first been released by AOL~\cite{pass2006picture} and MSN~\cite{zhang2006some}, containing thousands of search queries. Later on, Yandex\footnote{\url{https://www.yandex.com}} provided a dataset with $35$ million anonymized search sessions~\cite{serdyukov2014log}. Recently, Sogou\footnote{\url{https://www.sogou.com}} has made available a dataset of $537,\!000$ queries in Chinese, accompanied with $12.2$ million user interactions (Sogou-QCL)~\cite{zheng2018sogou}. Another recent IR collection in the web domain, MS~MARCO~\cite{nguyen2016ms}, provides a large set of informational question-style queries from Bing's search logs. These queries are accompanied by human-annotated relevant/non-relevant passages and documents. More recently, the ORCAS collection~\cite{craswell2020orcas} releases a large dataset of the click logs related to MS~MARCO.

In the health domain, several standard IR benchmarks have been developed over the years, especially through evaluation campaigns such as the Text Retrieval Conference (TREC) and Conference and Labs of the Evaluation Forum (CLEF). Examples of some IR tasks are CLEF eHealth Consumer Health Search~\cite{jimmy2018overview} and TREC Precision Medicine~\cite{Roberts2019OverviewOT}. The related collections consists of some dozens of queries, where each query is accompanied by a set of human-annotated relevance judgements on documents. \coll complements the previous efforts in creating standard health IR collections, by providing a novel dataset of health queries and query-document relevance signals, several orders of magnitude larger in size.

\label{sec:related}

\section{\coll Logs Dataset}
\label{sec:dataset}

\begin{table}[t]
\begin{center}
\caption{Statistics of the \coll logs dataset.} 
\begin{tabular}{l r}
\toprule
Number of click log entries & 5,272,064\\
Number of sessions & 1,602,648\\
Average number of q-d interactions per session & 3.3\\
Number of unique queries & 1,647,749\\
Number of documents (clicked or retrieved) & 2,347,977\\

\bottomrule
\end{tabular}
\label{tbl:logdataset} 
\end{center}
\end{table}

The \coll logs dataset consists of the user interactions of the Trip search engine collected between January 2013 and October 2020. A sample click log entry is shown in Figure~\ref{fig:logs}. Each entry consists of date and time of search (in Unix time, in milliseconds), search session identifier, submitted query (\texttt{Keywords} field), document identifiers of the top 20 retrieved documents,\footnote{The top 20 retrieved documents by the search engine are shown to users in one page. The retrieved documents are only available in the log files since August 2016} and the metadata of the clicked document. For the clicked document, the provided data contains its unique identifier and URL. If the clicked document is a scientific publication, its title, DOI, and clinical areas are also stored. We should emphasize that the privacy of individual users is preserved in the clicked search logs by cautiously removing any Personally Identifiable Information, (PII).

\begin{figure}[t]
    \centering
    \includegraphics[width=0.32\textwidth]{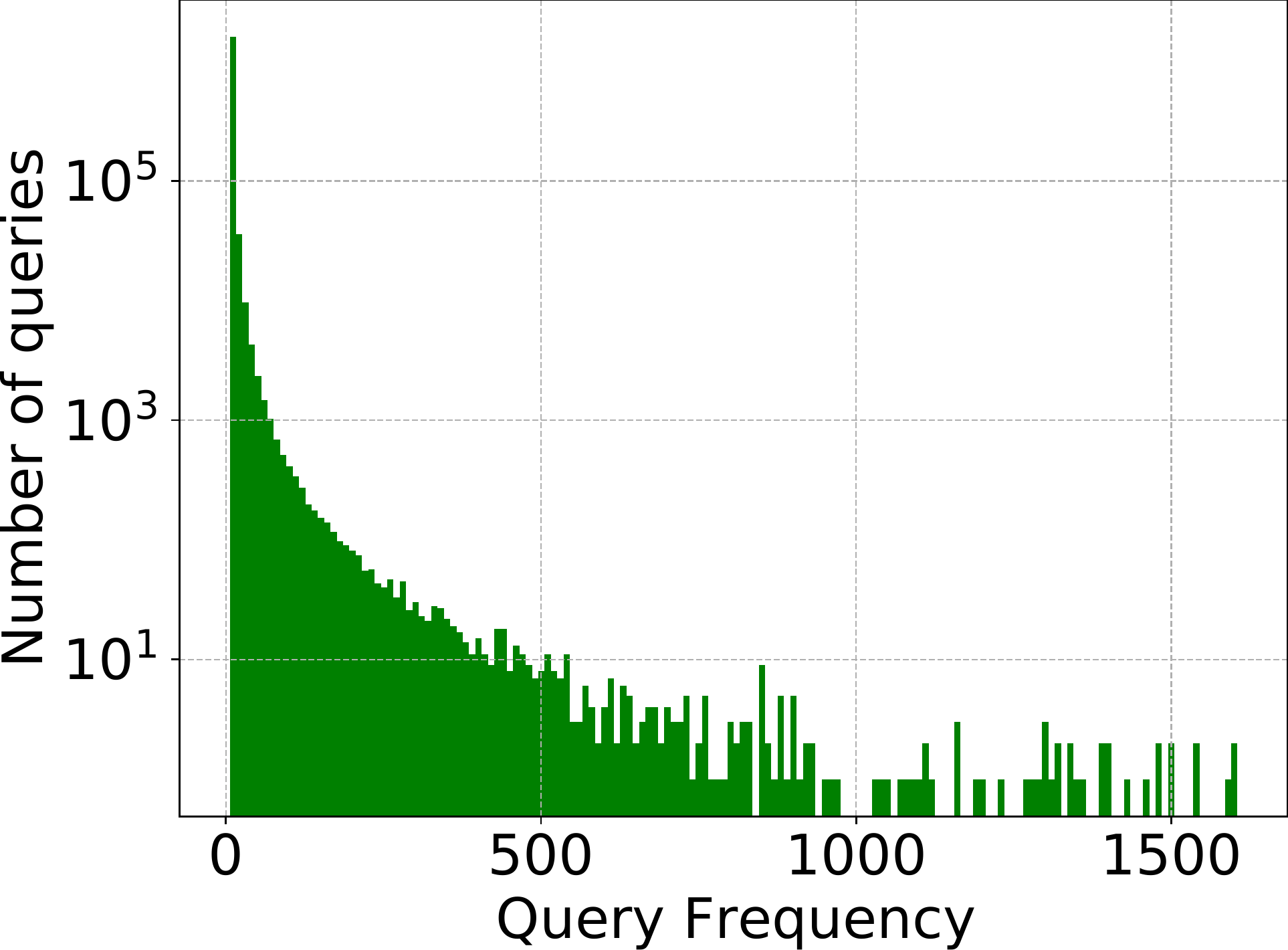}
    \caption{Query frequency histogram. The vertical axis is presented in log scale.}
    \label{fig:qry_hist}
\end{figure}

The statistics of the \coll logs dataset are reported in Table~\ref{tbl:logdataset}. It consists of approximately $5.2$ million click log entries, appeared in around $1.6$ million search sessions ($\sim\!3.3$ interactions per session). 

The click logs contain around $1.6$ million unique queries. These queries appear in the logs at varying frequencies. Figure~\ref{fig:qry_hist} shows the log-scaled query frequency histogram. The histogram follows an exponential trend -- there are many rare queries (issued only a few times to the search engine), while there are few highly frequent ones. Examples of a frequent and a rare query are \emph{``asthma pregnancy''}, and \emph{``antimicrobial activity of medicinal plants''}, respectively.

As reported in Table~\ref{tbl:logdataset}, the log files contain approximately $2.3$ million documents. Together with the dataset of click logs, we provide the corresponding titles and URLs of all documents. Examining the origin of clicked documents, we observe that approximately $80\%$ of the documents point to articles in the MEDLINE catalog, around $11\%$  to entries in \url{https://clinicaltrials.gov}, and the rest to various publicly available resources on the web.  

\begin{figure}[t]
    \centering
    \includegraphics[width=0.32\textwidth]{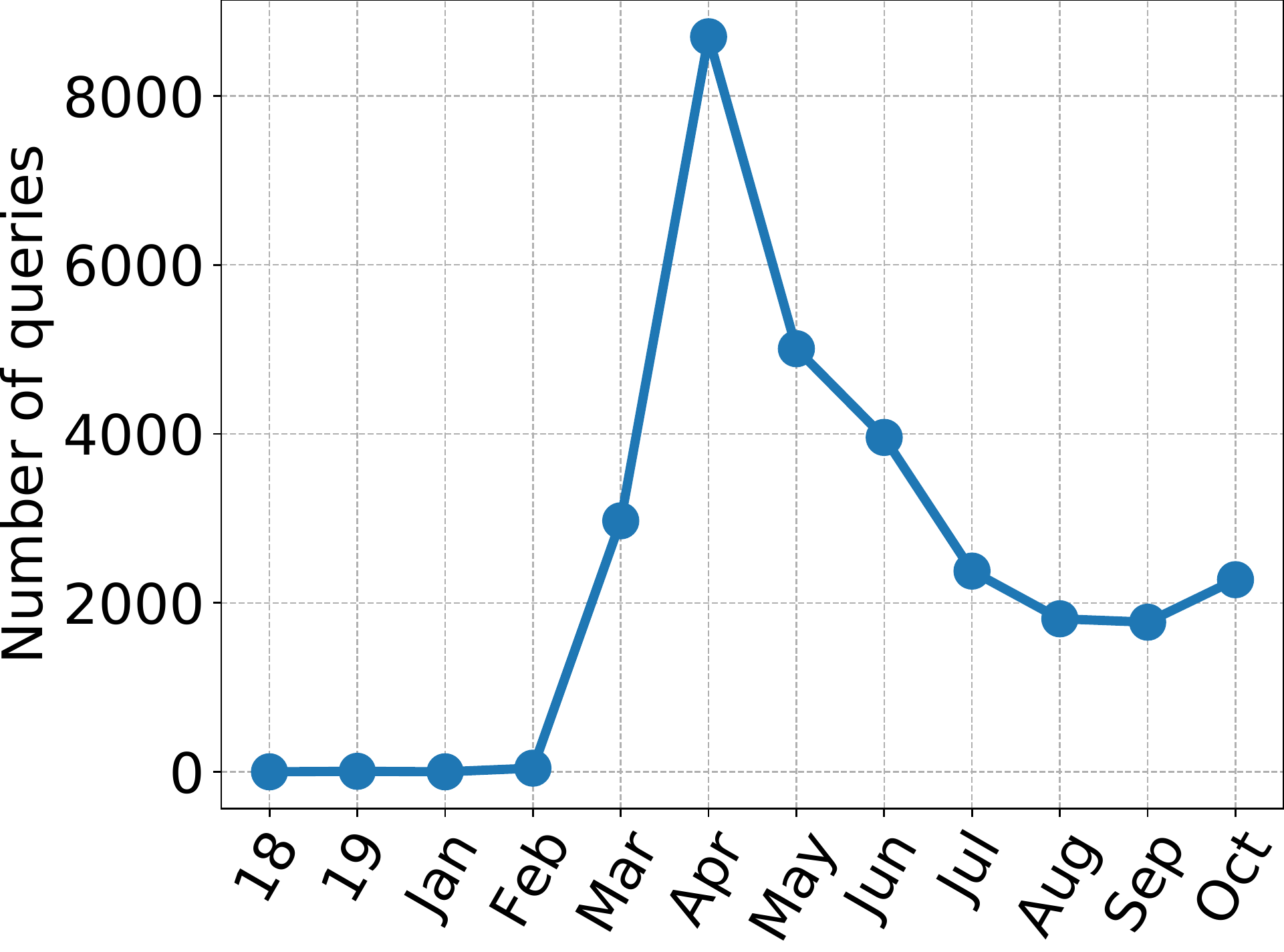}
    \caption{Number of submitted queries related to the COVID-19 pandemic. The entries 2018 and 2019 compound all occurrences in those years.}
    \label{fig:qry_corona_count}
\end{figure}

Finally, looking at the query contents, Figure~\ref{fig:qry_corona_count} reports the number of times a query related to the COVID-19 virus\footnote{We count those queries containing the keywords \emph{corona}, \emph{covid}, \emph{covid-19}, and \emph{covid19}} is submitted to the search engine in the period of 2018-2020. The data for 2018 and 2019 are presented as annual sums, while for the year 2020, numbers are reported per month. While there are only few COVID-19-related queries before the February of 2020, the information need rapidly gains popularity with a peak in April. The provided data is potentially a useful resource for studying the COVID-19 pandemic, as well as the reaction and evolution of search engines regarding the sudden emergence of previously unknown/uncommon diseases.

\section{\coll Health IR Benchmark}
\label{sec:benchmark}

\begin{table*}[t]
\begin{center}
\caption{Statistics of \coll IR benchmark.} 
\begin{tabular}{l r}
\toprule
Number of query-document interactions & 4,054,593\\
Number of documents & 1,523,878\\
Number of queries (all/\head/\torso/\tail) & 692,699 / 5,879 / 108,314 / 578,506\\
Average query length & $4.4\pm 2.4$ \\ 
Average document length & $259.0\pm 81.7$\\
\cdashlinelr{1-2}

Number of relevance data points in \raw (all/\head/\torso/\tail) & 2,870,826 / 246,754 / 994,529 / 1,629,543 \\
Average relevance data points per query in \raw (\head/\torso/\tail) & 41.9 / 9.1 / 2.8 \\
Number of relevance data points in \dctr (\head) & 263,175  \\ 
Average relevance data points per query in \dctr (\head) & 46.2 \\\cdashlinelr{1-2} 

Number of queries used to create training set & 685,649\\ 
Number of non-zero \raw relevance data points used to create training set & 1,105,811\\ 
Number of items in training set & 23,222,038\\
Number of queries in validation sets (\head/\torso/\tail) & 1,175 / 1,175 / 1,175\\
Number of queries in test sets (\head/\torso/\tail) & 1,175 / 1,175 / 1,175\\
\bottomrule
\end{tabular}
\label{tbl:benchmark} 
\end{center}
\end{table*}


To create the \coll benchmark, we use a subset of click log entries that refer to those documents that are indexed in the MEDLINE catalog. This choice was made because the majority of the click logs refer to MEDLINE articles ($\sim\!80\%$). Additionally, from a practical point of view, considering that the MEDLINE articles remain constant over time, the contents of the corresponding documents can be conveniently determined from the present MEDLINE catalog, despite the fact that each document in the logs is accessed at some historic timestamp. MEDLINE articles are similarly used in several other health IR benchmarks~\cite{roberts2017overview,roberts2018overview,Roberts2019OverviewOT}. This subset encompasses around $4$ million log entries. The statistics of the \coll benchmark are reported in Table~\ref{tbl:benchmark}. The process of creating the benchmark is explained in the following.

We create the collection of documents that appear in the subset of click logs, resulting in approximately $1.5$ million unique documents. For each document, we fetch the corresponding article from the MEDLINE catalog. Similar to the TREC Precision Medicine track~\cite{roberts2017overview,roberts2018overview,Roberts2019OverviewOT}, we use the title and abstract of the articles as documents of the \coll benchmark. 

We then extract the queries from the subset of click logs, resulting in around $692,\!000$ unique queries. As shown in Figure~\ref{fig:qry_hist}, many queries appear rarely while some few are submitted very often. In creating the benchmark, we are interested in the performance of various IR models on queries in different frequency ranges, namely the sets of infrequent, modestly-frequent, and highly-frequent queries. To this end, we split the queries into three groups, namely \head, \torso, and \tail, such that the queries in this sets cover 20\%, 30\%, and 50\% of the search engine traffic (according to the subset of click logs). This, in fact, results in assigning the queries with frequencies lower than 6 to \tail, the ones between 6 and 44 to \torso, and all the rest with frequencies higher higher than 44 to \head. The number of queries in each group is reported in the upper section of Table~\ref{tbl:benchmark}. While the numbers of unique queries in \head and \torso are much smaller than those in \tail, the former together still cover half of the search engine's traffic since their queries repeat much more often than the ones of \tail.

Next, we create two sets of query-to-document relevance signals, each created using a click-through model. The first relevance set, referred to as \raw, follows a simple approach by considering every clicked document as relevant to its corresponding query. The second set uses the Document Click-Through Rate (\dctr)~\cite{craswell2008experimental,chuklin2015click}. Creating two sets using different click-through models provides insight about the effect of each click-through model on the final evaluation results, achieved using the corresponding relevance signals.

To calculate the two sets of relevance scores, we first collect all retrieval information related to each query, consisting of the retrieved documents and the clicked ones. In the \raw set, for a given query, a relevance score of 1 is assigned to each of its clicked documents. For completeness, we also include a set of non-relevant documents (relevance score of 0) for each query, consisting of the documents in the ranked list of the query that appear in higher positions than the clicked one. This in fact follows the common assumption in click-through models, that the user has checked the documents in the retrieved ranked list from top till the clicked document, and has not found the top non-clicked ones relevant~\cite{chuklin2015click}. We should note that adding these non-relevant scores typically does not affect the evaluation results, as relevance scores of 0 are commonly ignored.

\begin{table*}[t]
\begin{center}
\caption{Evaluation results on the \coll benchmark using \raw relevance information. The best results for each metric are indicated by bold numbers. The superscript letters indicate significant improvements ($p<0.05$) over the other models, indicated with letters inside the parentheses: the superscript letter $a$ refers to BM25 and \prf, $b$ to \pacrr, $c$ to \matchpyramid, $d$ to \knrm, $e$ to \convknrm, and $f$ to \tk.} 
\begin{tabular}{l l l l  l l l}
\toprule
\multirow{2}{*}{Model}  & \multicolumn{3}{c}{Validation} & \multicolumn{3}{c}{Test} \\
& \multicolumn{1}{c}{NDCG} & \multicolumn{1}{c}{MRR} & \multicolumn{1}{c}{Recall} & \multicolumn{1}{c}{NDCG} & \multicolumn{1}{c}{MRR} & \multicolumn{1}{c}{Recall} \\\midrule

\multicolumn{7}{c}{\head}\\\midrule
BM25 ($a$) & $0.209$ & $0.362$ & $0.129$  & $0.199$ & $0.347$ & $0.128$ \\
\prf ($a$) & $0.205$ & $0.344$ & $0.129$ & $0.199$ & $0.354$ & $0.125$ \\
\pacrr ($b$) & $0.254^{a}$ & $0.451^{a}$ & $0.151^{a}$ & $0.234^{a}$ & $0.410^{a}$ & $0.142^{a}$ \\
\matchpyramid ($c$) & $0.275^{ab}$ & $0.479^{ab}$ & $0.160^{ab}$ & $0.244^{ab}$ & $0.419^{a}$ & $0.150^{ab}$ \\
\knrm ($d$) & $0.268^{ab}$ & $0.466^{a}$ & $0.156^{a}$ & $0.254^{abc}$ & $0.449^{abc}$ & $0.151^{ab}$ \\
\convknrm ($e$) & $0.279^{abd}$ & $0.490^{abd}$ & $0.159^{ab}$ & $0.266^{abcd}$ & $0.473^{abcd}$ & $0.152^{ab}$ \\
\tk ($f$) & $\textbf{0.302}^{abcde}$ & $\textbf{0.521}^{abcde}$ & $\textbf{0.174}^{abcde}$ & $\textbf{0.284}^{abcde}$ & $\textbf{0.487}^{abcd}$ & $\textbf{0.167}^{abcde}$ \\\midrule
\multicolumn{7}{c}{\torso}\\\midrule
BM25 ($a$) & $0.224$ & $0.318$ & $0.271$ & $0.206$ & $0.283$ & $0.262$ \\
\prf ($a$) & $0.207$ & $0.290$ & $0.255$ & $0.194$ & $0.261$ & $0.254$ \\
\pacrr ($b$) & $0.230^{a}$ & $0.333$ & $0.271$ & $0.212$ & $0.302^{a}$ & $0.262$ \\
\matchpyramid ($c$) & $0.253^{abd}$ & $0.364^{ab}$ & $0.296^{abd}$ & $0.243^{ab}$ & $0.347^{ab}$ & $0.297^{abd}$ \\
\knrm ($d$) & $0.242^{ab}$ & $0.348^{a}$ & $0.286^{ab}$ & $0.235^{ab}$ & $0.338^{ab}$ & $0.283^{ab}$ \\
\convknrm ($e$) & $0.248^{ab}$ & $0.360^{ab}$ & $0.292^{ab}$ & $0.243^{ab}$ & $0.358^{abd}$ & $0.288^{ab}$ \\
\tk ($f$) & $\textbf{0.281}^{abcde}$ & $\textbf{0.394}^{abcde}$ & $\textbf{0.326}^{abcde}$ & $\textbf{0.272}^{abcde}$ & $\textbf{0.381}^{abcde}$ & $\textbf{0.321}^{abcde}$ \\\midrule
\multicolumn{7}{c}{\tail}\\\midrule
BM25 ($a$) & $0.285$ & $0.277$ & $0.429$ & $0.267$ & $0.258$ & $0.409$ \\
\prf ($a$) & $0.240$ & $0.227$ & $0.392$ & $0.242$ & $0.227$ & $0.384$ \\
\pacrr ($b$) & $0.289$ & $0.283$ & $0.429$ & $0.267$ & $0.261$ & $0.409$ \\
\matchpyramid ($c$) & $0.294$ & $0.293^{a}$ & $0.429$ & $0.281^{abe}$ & $\textbf{0.280}^{abde}$ & $0.409$ \\
\knrm ($d$) & $0.289$ & $0.279$ & $0.429$ & $0.272$ & $0.265$ & $0.409$ \\
\convknrm ($e$) & $0.289$ & $0.282$ & $0.429$ & $0.271$ & $0.265$ & $0.409$ \\
\tk ($f$) & $\textbf{0.310}^{abde}$ & $\textbf{0.298}^{a}$ & $\textbf{0.471}^{abcde}$ & $\textbf{0.295}^{abde}$ & $0.279$ & $\textbf{0.459}^{abcde}$ \\

\bottomrule
\end{tabular}
\label{tbl:results_raw} 
\end{center}
\end{table*}

\begin{table*}[t]
\begin{center}
\caption{Evaluation results using \dctr relevance information. Notations as in Table~\ref{tbl:results_raw}.}
\begin{tabular}{l l l l  l l l}
\toprule
\multirow{2}{*}{Model}  & \multicolumn{3}{c}{Validation} & \multicolumn{3}{c}{Test} \\
& \multicolumn{1}{c}{NDCG} & \multicolumn{1}{c}{MRR} & \multicolumn{1}{c}{Recall} & \multicolumn{1}{c}{NDCG} & \multicolumn{1}{c}{MRR} & \multicolumn{1}{c}{Recall} \\\midrule

\multicolumn{7}{c}{\head}\\\midrule
BM25 ($a$) & $0.149$ & $0.314$ & $0.145$ & $0.140$ & $0.290$ & $0.138$ \\
\prf ($a$) & $0.145$ & $0.296$ & $0.143$ & $0.141$ & $0.300$ & $0.136$ \\
\pacrr ($b$) & $0.186^{a}$ & $0.390^{a}$ & $0.166^{a}$ & $0.175^{a}$ & $0.356^{a}$ & $0.162^{a}$ \\
\matchpyramid ($c$) & $0.202^{ab}$ & $0.416^{ab}$ & $0.181^{ab}$ & $0.183^{a}$ & $0.372^{a}$ & $0.173^{ab}$ \\
\knrm ($d$) & $0.196^{ab}$ & $0.407^{a}$ & $0.174^{ab}$ & $0.191^{abc}$ & $0.393^{ab}$ & $0.173^{ab}$ \\
\convknrm ($e$) & $0.206^{abd}$ & $0.429^{abd}$ & $0.180^{ab}$ & $0.198^{abcd}$ & $0.420^{abcd}$ & $0.178^{ab}$ \\
\tk ($f$) & $\textbf{0.221}^{abcde}$ & $\textbf{0.453}^{abcde}$ & $\textbf{0.194}^{abcde}$ & $\textbf{0.208}^{abcde}$ & $\textbf{0.434}^{abcd}$ & $\textbf{0.189}^{abcde}$ \\
\bottomrule
\end{tabular}
\label{tbl:results_dctr} 
\end{center}
\end{table*}

Regarding the \dctr set, the relevance score of a document for a given query is defined as the number of times that the document is clicked divided by the number of times the document is retrieved in the result lists of the query. These scores have a numeric range from 0 to~1. 
To be able to use these scores for retrieval evaluation, we need to discretize them to relevance grades. To this end, we follow a similar approach to the one in \citet{xiong2017end}. In particular, we project the \dctr scores to 4 relevance grades (0 to 3), where 0 is non-relevant and 3 is highly relevant. The \dctr scores are discretized to these grades by selecting thresholds such that the relevance grades follow a similar distribution as TREC Web Track 2009-2012 query-relevance data. The selected thresholds are 0.0, 0.04, 0.3, and 1, resulting in a distribution of 71.4\%, 19.7\%, 6.0\%, and 2.9\% of scores for grades 0 to 3, respectively. We should note that similar to \citet{xiong2017end}, the \dctr model is only calculated for \head queries. This is due to the fact that the \dctr method provides meaningful relevance signals from click logs only if the queries are sufficiently frequent. The statistics of the numbers of relevance data points as well as their averages per query, for each group are reported in the center section of Table~\ref{fig:qry_hist}. We should note that while we provide two click-through models, the log files can be indeed exploited in future studies for creating further and more advanced click-through models. 


The provided documents, queries, and relevance signals are well suited for training neural IR models or as an evaluation benchmark. To enable consistent and reproducible training and evaluation in future studies, we construct pre-defined validation and test sets as well as pair-wise training data. In particular, for each group (\head, \torso, and \tail), we create validation and test sets by randomly selecting $1,\!175$ queries from the pool of the queries in the corresponding group. To create the training data, we use the remaining queries of the three groups ($\sim\!685,\!000$), and their non-zero \raw relevance datat points ($\sim\!1.1$ million). We follow the pair-wise learning to rank method~\cite{liu2010learning}, where each data entry is a triple, consisting of a query, a relevant, and a non-relevant document. Similar to \citet{nguyen2016ms}, for each relevant query-document pair, we create 20 training triples, where the query and relevant document are taken from the given estimated relevance, and the non-relevant document is randomly sampled from the top $1,\!000$ results of a BM25 model. This results in training data with more than $23$ million data items, as reported in the lower section of Table~\ref{fig:qry_hist}. We would like to point out that, considering the relatively high number of relevance signals per query especially in the \head and \torso group, training data can also be created for list-wise learning-to-rank approaches.



\section{Retrieval Experiments on \coll Benchmark}
\label{sec:experiments}

In this section, we demonstrate the usefulness of the proposed dataset for model training and benchmarking, by reporting the performance of various IR models on the \coll benchmark collection. We first explain our experimental setup, followed by presenting and discussing the evaluation results.  

\subsection{Experiment Setup}

\paragraph{IR Models}
We conduct studies using several classical IR models as well as recent neural ones. As strong classical IR baselines, we use BM25~\cite{robertson2009probabilistic} as a widely used exact matching model, and the RM3 Pseudo Relevance Feedback (PRF) model~\cite{lv2009comparative,lavrenko2001relevance} as a strong query expansion baseline. In addition, we study the effectiveness of five recent neural IR models, namely Position Aware Convolutional Recurrent Relevance Matching (\pacrr)~\cite{hui2017pacrr}, Match Pyramid (\matchpyramid)~\cite{pang2016text}, Kernel-based Neural Ranking Model (\knrm)~\cite{xiong2017end}, Convolutional KNRM (\convknrm)~\cite{dai2018convolutional}, and Transformer-Kernel (\tk)~\cite{Hofstaetter2020_sigir}. These neural models are selected due to their strong performance on retrieval tasks as well as their diversity in terms of model architectures.

\paragraph{Evaluation} 
Performance evaluation is carried out in terms of Mean Reciprocal Ranks~\cite{voorhees1999trec} (MRR), Recall at a cutoff of 10, and Normalized Discounted Cumulative Gain (NDCG) at a cutoff of 10. Statistical significance tests are conducted using a two-sided paired $t$-test and significance is reported for $p<0.05$. The evaluation is performed using \texttt{trec\_eval}.\footnote{\url{https://github.com/usnistgov/trec_eval}}

\paragraph{Hyper-parameters and Training}
For classical IR models, we use the default hyper-parameters of the Anserini toolkit~\cite{yang2017anserini}. For neural IR models, we use pre-trained word2vec Skipgram~\cite{distributed2013mikolov} embeddings with 400 dimensions, trained on biomedical texts from the MEDLINE dataset.\footnote{\url{http://nlp.cs.aueb.gr/software.html}} In a preprocessing step, all documents are casefolded by projecting all characters to lower case. We remove numbers and punctuation (except periods), and apply tokenization using AllenNLP WordTokenize~\cite{gardner2018allennlp}. The vocabulary set is created by filtering those terms with collection frequencies lower than 5, resulting in 215,819 unique terms. We use the Adam optimizer~\cite{kingma2014adam} with learning rate 0.001, a maximum of 3 epochs, and early stopping. We use a batch size of 64. The maximum length of queries and documents is set to 20 and 300 tokens, respectively. For \knrm, \convknrm, and \tk, we set the number of kernels to $11$ in the range of $-1$ to $+1$ with a step size of $0.2$, and standard deviation of $0.1$. The dimension of the convolutional vectors in \convknrm is set to $400$. The \tk model consists of $2$ layers of Transformers~\cite{vaswani2017attention} with $2$ heads and intermediate vector size of $512$. In \matchpyramid, the number of convolution layers is set to $5$, each with kernel size $3\times3$ and 16 convolutional channels. The pre-trained word embeddings are updated during training. The threshold for selecting the top $n$ retrieved documents for re-ranking is chosen by tuning the $n$ parameter on a range from $1$ to $100$, based on the NDCG results of the validation set. More information about training and reproducing these baseline models as well as the results of other models is provided in the collection's web page: {\collurl}.

\subsection{Evaluation Results}
The evaluation results on the validation and test sets of \head, \torso, and \tail using \raw relevance information are shown in Table~\ref{tbl:results_raw}. Table~\ref{tbl:results_dctr} reports the evaluation results on the \head queries using the \dctr relevance information.\footnote{Please note that \dctr results are only meaningful for \head (see Section~\ref{sec:benchmark}).} The best results for each evaluation metric are shown in bold. Significant improvements over the other models are indicated with the superscript letters inside the parentheses in front of the models. For brevity, we assign the same sign of significance to the two classical baselines (superscript letter $a$), indicating significant improvements over both models. 

In general, the neural models significantly outperform the classical ones, where the \tk model in particular shows the best overall performance by significantly outperforming the classical IR models across all groups and evaluation metrics. We observe similar patterns between the results of \dctr and \raw on the \head set. The overall achieved improvements with neural models are more prominent for groups containing more frequent queries, namely the improvements of the queries in \head are higher than the ones in \torso, and subsequently in \tail.

The evaluation results on the \coll benchmark and specifically the improvements of the various neural models relative to each other are similar to the behavior observed on the MS~MARCO collection in previous studies~\cite{hofstatter2019effect,Hofstaetter2020_sigir}. This is in particular the case for the results of \head (according to both \raw and \dctr) and \torso groups. These results highlight the value of the provided benchmark and training data for research on neural and deep learning-based IR models in general, and in the health domain in specific.

\section{Conclusion}
This work provides a novel click-log dataset covering the 7 years user interactions of a health search engine. The dataset consists of approximately $5.2$ million user interactions. Based on the dataset, we create \coll, a novel large-scale health IR benchmark with approximately $700,\!000$ queries and $2.8$ million query-document relevance signals. We use \coll to train several neural IR models and evaluate their performances on well-defined held-out sets of queries. The evaluation results in terms of NDCG, MRR, and Recall demonstrate the adequacy of \coll for training large, highly parametric IR models and show significant improvements of neural models over classical ones, particularly for queries that appear frequently in the log dataset. The log dataset as well as the created benchmark and training data are made available to the community to foster reproducible academic research on neural IR models, particularly in the health domain. 

\section*{Acknowledgements}
This research is supported in part by the NSF (IIS-1956221). The views and conclusions contained herein are those of the authors and should not be interpreted as necessarily representing the official policies, either expressed or implied, of NSF or the U.S. Government. Thanks to Zhuyun Dai for her help and advice on designing click-through models.

\bibliographystyle{ACM-Reference-Format}
\bibliography{reference}

\end{document}